\documentclass{ws-ijmpc_rev}
\usepackage{graphicx}

\begin{document}

\markboth{H. Koibuchi and A.Shobukhov}
{Internal phase transition induced by external forces}

\catchline{}{}{}{}{}

\title{Internal phase transition induced by external forces in Finsler geometric model for membranes
}

\author{Hiroshi Koibuchi$^{1}$\, and Andrey Shobukhov$^{2}$\
}

\address{$^{1}$\ Department of Mechanical and Systems Engineering, National Institute of Technology, Ibaraki College, Nakane 866 Hitachinaka, Ibaraki 312-8508, Japan
\\koibuchi@mech.ibaraki-ct.ac.jp
}
\address{$^{2}$\ Faculty of Computational Mathematics and Cybernetics, Lomonosov Moscow State University, \\
119991, Moscow, Leninskiye Gory, MSU, 2-nd Educational Building, Russia
}

\maketitle

\begin{history}
\received{2015}
\revised{2015}
\end{history}

\begin{abstract}
We numerically study an anisotropic shape transformation of membranes under external forces for two-dimensional triangulated surfaces on the basis of Finsler geometry. The Finsler metric is defined by using a vector field, which is the tangential component of a three dimensional unit vector $\sigma$ corresponding to the tilt or some external macromolecules on the surface of disk topology. The sigma model Hamiltonian is assumed for the tangential component of $\sigma$ with the interaction coefficient $\lambda$.  For large (small) $\lambda$, the surface becomes oblong (collapsed) at relatively small bending rigidity. For the intermediate $\lambda$, the surface becomes planar. Conversely, fixing the surface with the boundary of area $A$ or with the two point boundaries of distance $L$, we find that the variable $\sigma$ changes from random to aligned state with increasing of $A$ or $L$ for the intermediate region of $\lambda$. This implies that an internal phase transition for $\sigma$ is triggered not only by the thermal fluctuations but also by external mechanical forces. We also find that the frame (string) tension shows the expected scaling behavior with respect to $A/N$ ($L/N$) at the intermediate region of $A$ ($L$) where the $\sigma$ configuration changes between the disordered and ordered phases. Moreover, we find that the string tension $\gamma$ at sufficiently large $\lambda$ is considerably smaller than that at small $\lambda$. This phenomenon resembles the so-called soft-elasticity in the liquid crystal elastomer, which is deformed by small external tensile forces. 
\end{abstract}

\ccode{PACS Nos.: 11.25.-w,  64.60.-i, 68.60.-p, 87.10.-e, 87.15.ak}

\section{Introduction}\label{intro}
A considerable number of studies have been conducted on the origins of anisotropic shape transformation (AST) in membranes.  The AST is a typical morphological change in membranes, and almost all morphological changes are regarded as ASTs.  As an example of AST, the prolate shape of fluid vesicles has its origin in the negative pressure ${\it \Delta}p$ inside the surface \cite{GOMPPER-KROLL-SMMS2004,GomKro-PRE1995}. The flow field is also considered to be the origin of AST for biological membranes such as red blood cells \cite{KWSLipow-PRL1996,Nog-Gom-PRL2004,KS-PRL2006}. Fluid membranes are expected to undergo AST when they are supported by the cytoskeleton  even if it is isotropic \cite{Koibuchi-PRE2007}. Also for better understanding of the ASTs the bilayer area difference elasticity (ADE) model is proposed \cite{Sefert-etal-PRA1991,Sefert-etal-PRE1994}. Together with the line tension energy, the ADE Hamiltonian is used for the analysis of experimentally observed ASTs such as the circular-to-stripe domains transition in the two-component membrane \cite{Yanagisawa-Imai-Tani-PRL2008,Yanagisawa-Imai-Tani-PRE2010}. Moreover, the originating of the AST from the anisotropic bending rigidity has been studied extensively using the  Landau-Ginzburg model for membranes \cite{Radzihovsky-SMMS2004,Radzihovsky-Toner-PRL1995,Radzihovsky-Toner-PRE1998}. 

The three-dimensional structure of the lipids (tilts) or of some other molecules is also considered as the origin of AST \cite{Koibuchi-Sekino-PhysicaA2014}. The internal molecular structure influences the external macroscopic structure of the material. In other words, the AST is naturally understood as a phenomenon induced by the internal phase transition in the context of Finsler geometry. 

However, there is another question: whether the converse phenomena occur or not. The problem is whether or not the internal order/disorder transition of tilts is activated by AST. If this internal transition is activated, it implies that an external mechanical force can influence the internal molecular structure via the surface transformation, which is not only AST but also isotropic shape transformations. We should note that the change in the alignment of the liquid crystal molecules due to the external macroscopic force is considered as the origin of the so-called soft-elasticity in the liquid crystal elastomer (LCE), the elongation of which occurs with small external force \cite{Domenici-PNMRS-2012}.

This paper aims at seeing whether or not external mechanical forces cause the internal order/disorder transition of $\sigma$ in a Finsler geometric model, which is an extension of Helfrich and Polyakov (HP) model for membranes  \cite{HELFRICH-1973,POLYAKOV-NPB1986,FDAVID-SMMS2004}. The external force, which is the frame tension, is supplied for fixing the surface with the boundary vertices \cite{Cai-Lub-PNelson-JFrance1994}. The projected area $A_P$ inside this boundary is fixed by the frame tension. When $A_P$ is sufficiently large, the variable $\sigma$ is expected to be aligned to a spontaneously chosen direction. Another external force is also supplied to fix a surface with two vertices separated by a distance $L$ on the diagonal line of the surface. This force is identified with the string tension $\gamma$ \cite{Ambjorn-NPB1993,WHEATER-JP1994}. The surface becomes oblong with sufficiently large $L$, and in this case the variable $\sigma$ is expected to align along the oblong direction. The variable $\sigma$ strongly aligns to this direction when the coefficient $\lambda$ of the sigma model Hamiltonian is sufficiently large, and it is expected that $\gamma$ becomes very small in this case. This phenomenon resembles the soft-elasticity observed in the LCE as mentioned above. The LCE is a three-dimensional object and hence it is quite different from membranes, however, we expect the same mechanism works also for the two-dimensional case. In this paper, we will confirm that these expected phenomena are observed in the Finsler geometric surface model.

\section{Model}\label{model}
\begin{figure}[hbt]
\centering
\includegraphics[width=8.5cm]{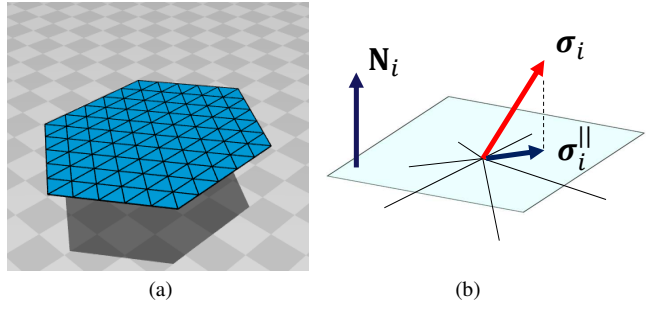}  
\caption{(a) A triangulated disk of size $(N,N_E,N_P)\!=\!(91,240,150)$.  (b) The tangential plane at the vertex $i$, and its unit normal vector ${\bf N}_i$. The variable $\sigma_i(\in S^2) $ and its tangential component $\sigma_i^{||}$. }
\label{fig-1}
\end{figure}
The discrete model is defined on the disk. By dividing the edges of the hexagon into $\ell$ pieces, we have a hexagonal lattice of size
\begin{eqnarray}
\label{lattice_size}
(N,N_E,N_P)=(3\ell^2+3\ell+1, 9\ell^2+3\ell, 6\ell^2), 
\end{eqnarray}
where $N,N_E$ and $N_P$ are the total number of vertices, the total number of edges, and the total number of plaquettes (or triangles). The disk for $\ell\!=\!5$ is shown in Fig. \ref{fig-1}(a).  The unit normal vector ${\bf N}_i$ at the vertex $i$ is given by 
\begin{eqnarray}
\label{unit-normal-vect}
{\bf N}_i = \sum_{j(i)}{\bf n}_{j(i)} A_{j(i)} / \mid \sum_{j(i)}{\bf n}_{j(i)} A_{j(i)}\mid,
\end{eqnarray}
where ${\bf n}_{j(i)}$ and $A_{j(i)}$ are respectively the unit normal vector and the area of the $j$-th triangle linked to the vertex $i$ \cite{KOIBUCHI-JPRE2004}. This ${\bf N}_i$ uniquely defines the tangential plane at the vertex $i$ (Fig. \ref{fig-1}(b)).

\begin{figure}[hbt]
\centering
\includegraphics[width=8.5cm]{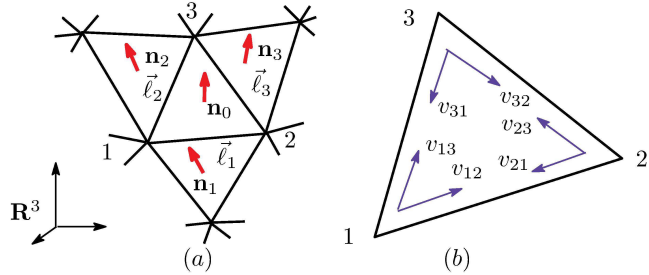}  
\caption{(a) The triangle $123$ and its three neighboring triangles, where ${\bf n}_i (i=0,..,3)$ are the unit normal vectors, and the edge vectors are given by $\vec\ell_1\!=\!{\bf r}_2\!-\!{\bf r}_1$, $\vec\ell_2\!=\!{\bf r}_3\!-\!{\bf r}_1$, $\vec\ell_3\!=\!{\bf r}_3\!-\!{\bf r}_2$. (b) The directions of the parameters $v_{ij}$.   }
\label{fig-2}
\end{figure}
The model is defined by the Hamiltonian
\begin{eqnarray}
\label{Disc-Eneg-1}
&&S=\lambda S_0+S_1+\kappa S_2,\quad S_0=-\sum_{ij} \sigma_i^{\mid\mid}\cdot \sigma_j^{\mid\mid}, \nonumber \\
&&S_1=\frac{1}{6}\sum_{\it \Delta}S_1({\it \Delta}), \quad S_2=\frac{1}{6}\sum_{\it \Delta}S_2({\it \Delta}),
\end{eqnarray}
where the variable $\sigma_i^{\mid\mid}$ is the tangential component of $\sigma_i$ (Fig. \ref{fig-1}(b)), and $\sum_{ij}$ in $S_0$ is the sum over all nearest neighbor vertices $ij$ linked by a bond, which is an edge of the triangles. We should note that $\sigma_i$ has a value on the unit sphere ($\sigma_i\in S^2$) and $\sigma_i^{\mid\mid}$ is given by 
\begin{eqnarray}
\label{sigma-parallel}
\sigma_i^{\mid\mid}=\sigma_i-(\sigma_i\cdot {\bf N}_i){\bf N}_i.
\end{eqnarray}

Since $\sigma_i^{\mid\mid}$ does not always belong to $S^2$, $S_0$ in Eq. (\ref{Disc-Eneg-1}) is different from the ordinary sigma model Hamiltonian.  
 $S_1({\it \Delta})$ and $S_2({\it \Delta})$ in Eq. (\ref{Disc-Eneg-1}) are given by 
\begin{eqnarray}
\label{Disc-Eneg-2} 
&&S_1({\it \Delta})= \gamma_{1}\ell_1^2+\gamma_{2}\ell_2^2+\gamma_{3}\ell_3^2, \nonumber \\
&&S_2({\it \Delta})=\kappa_{1}(1-{\bf n}_0\cdot{\bf n}_1)  +\kappa_{2}(1-{\bf n}_0\cdot{\bf n}_2) +\kappa_{3}(1-{\bf n}_0\cdot{\bf n}_3), 
\end{eqnarray} 
where the coefficients $\gamma_{i}$ and $\kappa_{i}$ are defined by
\begin{eqnarray}
\label{coefficients-2}
&&\gamma_{1}=\frac{v_{13}}{v_{12}}+\frac{v_{12}}{v_{13}}, \;\gamma_{2}=\frac{v_{32}}{v_{31}}+\frac{v_{31}}{v_{32}}, \;\gamma_{3}=\frac{v_{21}}{v_{23}}+\frac{v_{23}}{v_{21}}, \nonumber \\
&&\kappa_{1}=\frac{v_{13}}{v_{12}}+\frac{v_{23}}{v_{21}}, \;\kappa_{2}=\frac{v_{12}}{v_{13}}+\frac{v_{32}}{v_{31}}, \;\kappa_{3}=\frac{v_{21}}{v_{23}}+\frac{v_{31}}{v_{32}}. 
\end{eqnarray}
The parameters $v_{ij}$ are given by
\begin{equation}
\label{anisotropic-v} 
v_{ij}= 1+\left[\sigma_{ij}\right],\quad \sigma_{ij}=N_v \left|\sigma_i\cdot {\bf t}_{ij}\right|, 
\end{equation} 
where $[x]$ stands for ${\rm Max}\left\{n\in {\bf Z}|n\leq x\right\}$, $N_v\!=\!100$, and ${\bf t}_{ij}$ is the unit tangential vector from the vertex $i$ to the vertex $j$. The inner product $\sigma_i\cdot {\bf t}_{ij}$ in Eq. (\ref{anisotropic-v}) can also be written as $\sigma_i^{\mid\mid}\cdot {\bf t}_{ij}$. We should note that $v_{ij}$ in Eq. (\ref{anisotropic-v}) can also be written as $v_{ij}\!=\!|\sigma_i\cdot {\bf t}_{ij}| +\epsilon\; (\epsilon\!=\!0.01)$, which is almost (but not exactly) equivalent with the original one. This correspondence comes from the fact that $v_{ij}$ always appears as a rational form  $v_{ij}v_{kl}^{-1}$ in the Hamiltonians. Note also that the cut off $\epsilon$ is introduced to protect $v_{ij}v_{kl}^{-1}$ from being divergent. The cut off $\epsilon$ effectively makes the variable $\sigma_i$ not perpendicular to the direction ${\bf t}_{ij}$. 

We should note that the coefficients $\gamma_i$ and $\kappa_i$ are defined on the edges of triangles. On the boundary edges the coefficients are determined only by the triangle in one side of the edge, while on the internal edges they are determined by the two neighboring triangles. Note also that $\kappa_i$ at the bond $i$ becomes small (large) when $\sigma$ aligns parallel (perpendicular) to the direction of the bond $i$.   

The partition function is given by
\begin{eqnarray} 
\label{Part-Func}
 Z(\lambda,\kappa) =  \sum_{\sigma}\int \prod _i d {\bf r}_i \exp\left[-S(\sigma,{\bf r})\right],
\end{eqnarray} 
where the vertex number $i$ runs from $1$ to $N$. The center of mass of the surface is fixed for the free boundary surfaces. For the fixed boundary surfaces or the surfaces with fixed boundary points, $i$ in $ \prod _i$ runs only for the vertices which are not fixed. 

 In the limit of $\lambda\!\to\!\infty$, it is expected that the variable $\sigma$ has only the tangential component and hence  $\sigma_i\cdot {\bf N}_i\!=\!0$, which implies  $\sigma_i\!=\!\sigma_i^{\mid\mid}$. Moreover, the magnetization $M$ defined by 
\begin{eqnarray}
\label{magnetization} 
M=\sqrt{{\bf M}\cdot{\bf M}}, \quad {\bf M}=\sum _i \sigma_i
\end{eqnarray} 
becomes maximal in this limit. Thus, the energy $S_0$ effectively makes the surface stiffened (softened) in one direction (the other direction) at sufficiently large $\lambda$, and hence the surface is expected to be oblong if $\kappa$ is not too large. 

The discrete Hamiltonians $S_1$ and $S_2$ in Eq. (\ref{Disc-Eneg-1}) are obtained from the continuous Hamiltonians 
\begin{eqnarray}
\label{cont_S}
&&S_1=\int \sqrt{g}d^2x g^{ab} \partial_a {\bf r}\cdot \partial_b {\bf r}, \nonumber \\
&&S_2=\frac{1}{2}\int \sqrt{g}d^2x  g^{ab} \partial_a {\bf n} \cdot \partial_b {\bf n},  
\end{eqnarray} 
by replacing the metric $g_{ab}$ with the discrete Finsler metric $g_{ab}^F$ on a triangulated surface (Figs. \ref{fig-1}(a)--\ref{fig-1}(b)) such that \cite{Koibuchi-Sekino-PhysicaA2014}
\begin{equation}
\label{Finsler_metric}
g_{ab}^F=\left(  
       \begin{array}{@{\,}cc}
         v_{12}^{-2}  & 0\\
         0  & v_{13}^{-2}  
        \end{array} 
       \\ 
 \right).
\end{equation}
This $g_{ab}^F$ is obtained by deforming the diagonal elements of the Euclidean metric $g_{ab}\!=\!\delta_{ab}$ such that $1\!\to\! v_{12}^{-2}$ and $1\!\to\! v_{13}^{-2}$.
In $S_1$ of Eq. (\ref{cont_S}), the derivatives are replaced by $\partial_1 {\bf r} \!\to\! {\bf r}_2\!-\!{\bf r}_1$, $\partial_2 {\bf r} \!\to\! {\bf r}_3\!-\!{\bf r}_1$, where ${\bf r}_i$ denotes the position of the vertex $i$ such that $\ell_1\!=\!|{\bf r}_2\!-\!{\bf r}_1|$, $\ell_2\!=\!|{\bf r}_3\!-\!{\bf r}_1|$ (see Figs. \ref{fig-2}(a)--\ref{fig-2}(b)). The derivatives in $S_2$ can also be discretized in the following way: $\partial_1 {\bf n} \!\to\! {\bf n}_0\!-\!{\bf n}_2$, $\partial_2 {\bf n} \!\to\! {\bf n}_0\!-\!{\bf n}_1$, where ${\bf n}_i(i\!=\!0,1,2,3)$ are the unit normal vectors shown in Fig. \ref{fig-2}(a). The integration measure $\int \sqrt{g}d^2x$  and $g^{ab}$ in the continuous Hamiltonians are respectively replaced by $\int \sqrt{g}d^2x \!\to\! ({1}/{2})\sum_{\Delta} \sqrt{\det g_{ab}^F}$ and $g^{ab} \!\to\! \left(g^F\right)^{ab} \!=\! \left(g_{ab}^F\right)^{-1}$, where $g_{ab}^F$ is given by Eq. (\ref{Finsler_metric}).

\section{Simulation results}\label{results}
Firstly, we shall describe how to update the variables in MC simulations. The variable $\sigma (\in S^2)$ is randomly updated $\sigma\to\sigma^\prime\!=\!\sigma\!+\!\delta \sigma$ such that $\sigma^\prime \in S^2$, where the length of $\delta \sigma$ is chosen such that the acceptance rate becomes approximately $50\%\sim 70\%$. The vertex position is also updated randomly ${\bf r}\to{\bf r}^\prime\!+\!\delta {\bf r}$ so that the rate of acceptance is approximately $50\%$. One MC sweep (MCS) consists of $N$ sequential updates of $\sigma$ and $N$ sequential updates of $X$. 
\subsection{Surfaces with free boundary}\label{free-BND}
\begin{figure}[hbt]
\centering
\includegraphics[width=8.5cm]{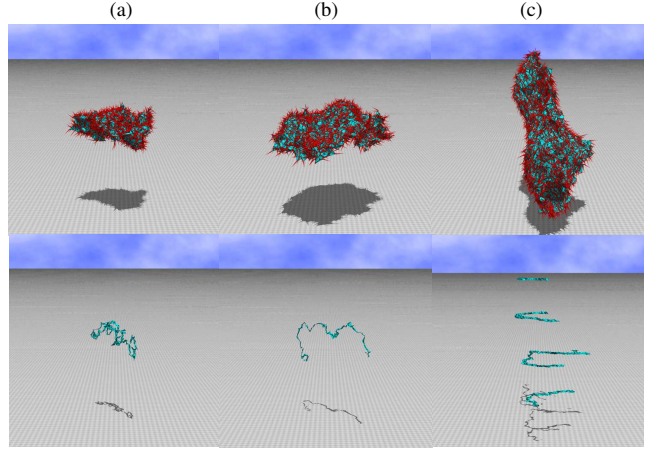}  
\caption{Snapshots of the surface and the surface section at (a) $\lambda\!=\!0.6$ (the collapsed phase), (b) $\lambda\!=\!0.76$ (the planar phase), and (c) $\lambda\!=\!1.0$ (the oblong phase). The bending rigidity is fixed to $\kappa\!=\!0.95$. The red burs represent the variables $\sigma_i$. The scales of the figures are the same.}
\label{fig-3} 
\end{figure}
In this subsection, we see the dependence of the phase structure on the parameter $\lambda$ under the fixed bending rigidity $\kappa$ of the model with the free boundary condition. As mentioned in Section \ref{model}, the surface is expected to be oblong (collapsed) at sufficiently large (small) $\lambda$ if $\kappa$ is an intermediate value. Indeed, we can see that the surface is collapsed at $\lambda\!=\!0.6$ while it becomes oblong at $\lambda\!=\!1$, and the planar phase appears between the collapsed and oblong phases when $\kappa\!=\!0.95$ (Figs. \ref{fig-3}(a)--(c)). Thus, these three different phases are expected to appear with varying $\lambda$ for $\kappa\!=\!0.95$. The phase boundary between the collapsed and planar phases is expected in the region $0.6<\lambda< 0.76$, and the phase boundary between the planar and oblong phases is expected in the region $0.76<\lambda<1$. We should note that the three phases can be seen even when $\kappa$ takes the value $\kappa\!=\!0.9$ or  $\kappa\!=\!1$. 

\begin{figure}[hbt]
\centering
\includegraphics[width=8.5cm]{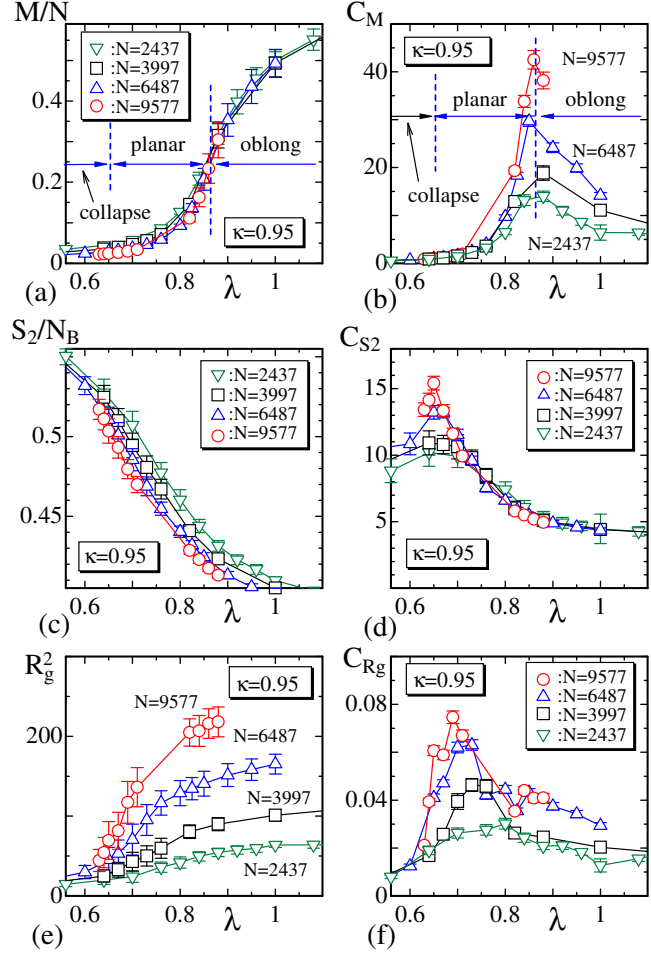}  
\caption{(a), (b) The magnetization $M/N$ and its variance $C_M$, (c), (d) the bending energy $S_2/N_B$ and the specific heat $C_{S_2}$, and (e), (f) the mean radius of gyration $R_g^2$ and its variance $C_{R^2}$. The bending rigidity is fixed to $\kappa\!=\!0.95$.}
\label{fig-4}
\end{figure}
To see the phase boundaries more clearly, the magnetization $M/N$ and the variance $C_M$, which is the magnetic susceptibility, defined by
\begin{eqnarray}
\label{CM}
C_M= \frac{1}{N}\langle\left( M-\langle M\rangle\right)^2\rangle
\end{eqnarray}
are plotted in Figs. \ref{fig-4}(a),(b). We find a peak in $C_M$ at $\lambda\simeq 0.86$, which is the boundary between the planar and oblong phases. This implies that the internal transition of the variable $\sigma$ induces the shape transformation between these two phases; the disordered (ordered) $\sigma$ makes the surface planar (oblong). To the contrary, we find no change in the configuration of $\sigma$ at the phase boundary, where $\lambda\simeq 0.65$, between the collapsed and planar phases. At $\lambda\geq 0.65$ the magnetization $M/N$ starts to grow (Fig. \ref{fig-4}(a)). The bending energy $S_2/N_B$ and the specific heat $C_{S_2}$ defined by   
\begin{eqnarray}
\label{CS2}
C_{S_2}= \frac{\kappa^2}{N}\langle\left( S_2-\langle S_2\rangle\right)^2\rangle
\end{eqnarray}
are plotted in Figs. \ref{fig-4}(c),(d). Contrary to the case of $M$, we see a peak in $C_{S_2}$ at the boundary ($\lambda\simeq 0.65$) between the collapsed and planar phases. The peak is expected to grow larger with increasing $N$, and therefore the transition is expected to be a second order one. The configurational change in  $\sigma$ causes this shape transformation transition, although the configuration of $\sigma$ changes continuously at this phase boundary. The mean square radius of gyration $R_g^2$ defined by
\begin{eqnarray}
\label{RG2}
R_g^2= \frac{1}{N}\sum_i \left({\bf r}_i-\bar{\bf r}\right)^2, \quad \bar{\bf r}=\frac{1}{N}\sum_i{\bf r}_i 
\end{eqnarray}
and the variance $C_{R^2}\!=\!(1/N)\langle\left( R_g^2-\langle R_g^2\rangle\right)^2\rangle$ are plotted in Figs. \ref{fig-4}(e),(f). The peak of $C_{R^2}$ at the phase boundary $\lambda\simeq 0.65$ reflects the shape transformation transition. This implies that the surface is effectively stiffened as $\lambda$ increases, although $\sigma$ appears to remain unchanged  at $\lambda\simeq 0.65$ from Figs. \ref{fig-4}(a),(b).

Here we comment on the reason why the phase transitions are reflected only in certain specific quantities. The quantities $S_2$ and $R_g^2$ rapidly change only at  $\lambda\simeq 0.65$, which is the boundary between the collapsed and planar phases, and these quantities change almost smoothly at $\lambda\simeq 0.86$, which is the boundary between the planar and oblong phases, although $C_{R^2}$ has a small peak at this boundary.  We should note that the oblong phase is the one and only anisotropic phase while the other two phases, the planar and collapsed phases, are isotropic. Note also that the surface strength, such as the surface tension and bending rigidity, along the spontaneously chosen direction becomes different from that along the other direction on the surface. From this property it follows that the anisotropic (isotropic) surface shape emerges as a result of the appearance of the ordered (disordered) phase of $\sigma$ in the context of Finsler geometric model. This is the reason why the magnetization $M/N$ rapidly varies and hence the variance $C_M$ has the peak only at $\lambda\simeq 0.86$, which is the phase boundary between the oblong and planar phases. On the other hand, the fact that the transition at $\lambda\simeq 0.65$ is reflected in $S_2$ and $R_g^2$ comes from the fact that the random configuration of $\sigma$ effectively softens the bending rigidity. This is another property of Finsler geometric model. This is possible  because the surface tension and the bending rigidity effectively changes according to Eq. (\ref{coefficients-2}), however, we don't go into detail any further on this point.

\begin{figure}[hbt]
\centering
\includegraphics[width=12.5cm]{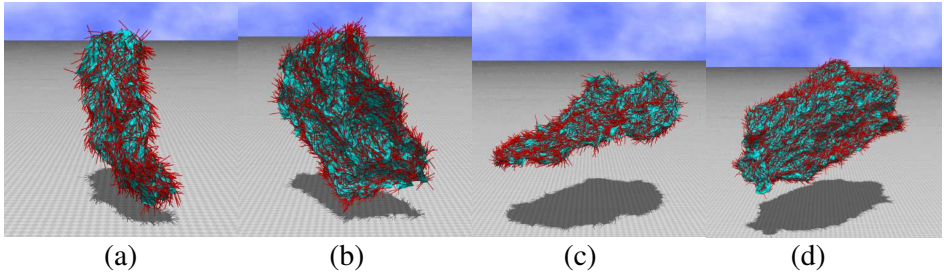}  
\caption{Snapshots of surfaces of size (a) $N\!=\!2437$, (b) $N\!=\!3997$, (c) $N\!=\!6487$, and (c) $N\!=\!9577$, obtained at $\lambda\!=\!1.1$ for $\kappa\!=\!0.95$. The variable $\sigma$, which is shown by red bur,  aligns along a spontaneously chosen direction on the surfaces. }
\label{fig-5}
\end{figure}

In Fig. \ref{fig-5} we show the snapshots obtained at $\lambda\!=\!1.1$ for $\kappa\!=\!0.95$ to see the shape of surface in the oblong phase. We actually find that the surface is almost flat but it is oblong in a spontaneously chosen direction.

\begin{figure}[hbt]
\centering
\includegraphics[width=8.5cm]{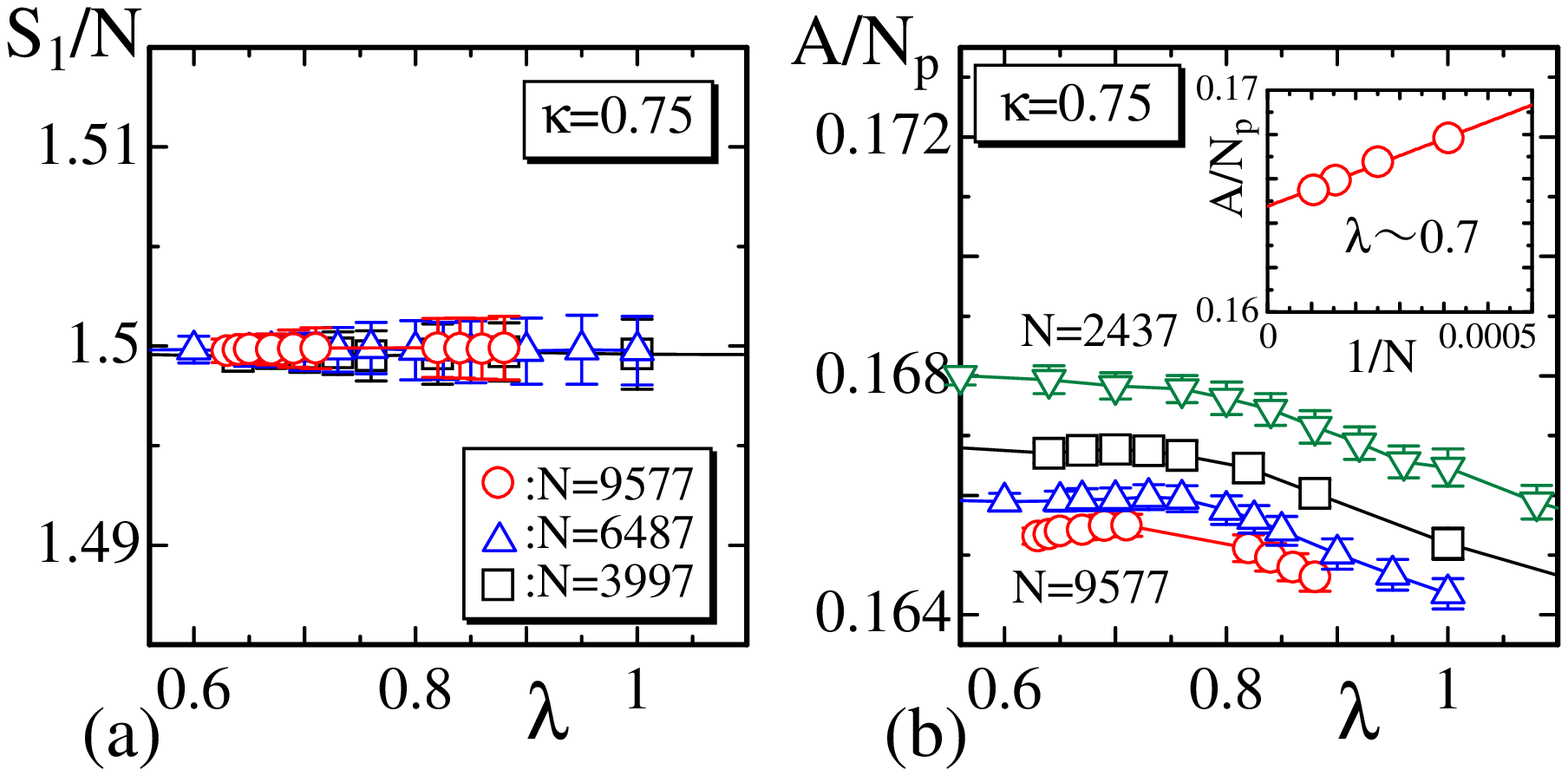}  
\caption{(a) The Gaussian bond potential $S_1/N$ vs. $\lambda$, and (b) the triangle area $A/N_P$ vs. $\lambda$.}
\label{fig-6}
\end{figure}
Finally in this subsection, we check that $S_1/N\!=\!3/2$ is satisfied. This relation is a direct consequence of the scale invariance of the partition function for the surface without the boundary condition except that the center of mass of the surface is fixed. We find from Fig. \ref{fig-6}(a) that the expected relation $S_1/N\!=\!3/2$ is satisfied. This implies that the simulations are successfully performed. 

The mean triangle area $A/N_P$ plotted in Fig. \ref{fig-6}(b) is slightly dependent on both $\lambda$ and $N$. This dependence of $A/N_P$ on $\lambda$ characterizes the Finsler geometric model, although the deviation in $A/N_P$ is very small. To the contrary, the triangle area is always constant in the canonical surface model, which is defined by $S_1\!=\!\sum_{ij}\ell_{ij}^2$ and $S_2\!=\!\sum_{ij}(1\!-\!{\bf n}_i\cdot{\bf n}_j)$.  In this canonical surface model, the relation  $S_1/N\!=\!3/2$ automatically implies that $A/N_P$ remains constant, because the regular triangles always dominate the ensemble configurations. 

An important feature of the model considered in this paper - that $A/N_P$ is not constant in the model of this paper - comes from the fact that the surface tension coefficients $\gamma_i$, $(i\!=\!1,2,3)$ in Eq. (\ref{coefficients-2}) are not constant and vary depending on the position and the direction on the surface. As a consequence, the bond length squares $\ell_i^2$, $(i\!=\!1,2,3)$, and hence the triangle area becomes dependent on the position and the direction on the surface. 
The dependence of $A/N_P$ on $N$ indicates that $\gamma_i$ or $v_{ij}$ in Eq. (\ref{anisotropic-v})  are strongly influenced by the size effect of the surface size $N$. Indeed, the linear plot of $A/N_P$ vs. $1/N$ at $\lambda\!\simeq\!0.7$ shown inside Fig. \ref{fig-6}(b) indicates that $A/N_P$ has a finite value in the limit of $N\!\to\!\infty$. This size effect implies that the long wave length modes in the fluctuations of $\sigma$ are important for the correlation energy between $\sigma^{||}$ in $S_0$. 

\subsection{Frame tension of surfaces with fixed boundary}\label{fixed-BND}
In this subsection, we see how the shape transformation is reflected in the internal degree of freedom $\sigma$. To see this, all boundary vertices are fixed on a plane so that the projected area $A_P$ enclosed by the boundary becomes a constant $A_P(=\!aN_P)$, where 
\begin{eqnarray}
\label{mean_p_area}
a=A_P / N_P
\end{eqnarray}
is the area of the projected triangle. It doesn't depend  on the lattice size $N$. This implies that the length of the diagonal line of the lattice of size $N$, such as shown in Fig. \ref{fig-1}(a), is given by $L\!=\!2(a/\sqrt{3})^{1/2}\ell$, where $\ell$ is the partition number of the hexagon edge introduced at the beginning of Section \ref{model}. This constant area $a$ is varied while $\lambda$ and $\kappa$ are fixed, and the dependence of the $\sigma$ configuration on $a$ is checked. We should note that $A_P$ is not always identical with the real surface area, which varies due to the surface fluctuations.

The boundary condition assumed in this subsection is used to calculate the frame tension $\tau$. As a consequence, we will see how the external force $\tau$ influences the internal variable $\sigma$.

\begin{figure}[hbt]
\centering
\includegraphics[width=8.5cm]{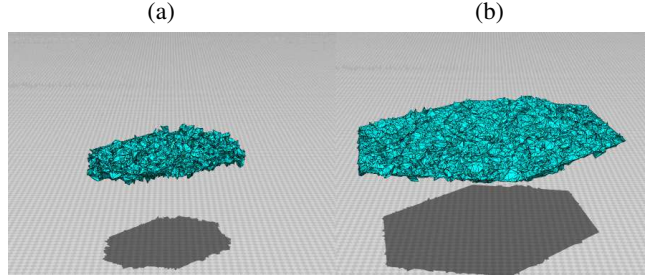}  
\caption{Snapshot of the surfaces of $N\!=\!9577$ for (a) $a\!\simeq\!0.027$ and (b) $a\!\simeq\!0.086$, where $\lambda\!=\!0.9$ and $\kappa\!=\!0.65$.  The variable $\sigma$ is not shown on the surfaces. }
\label{fig-7}
\end{figure}
Figures \ref{fig-7}(a),(b) show snapshots of the surfaces of size $N\!=\!9577$ for $a\!\simeq\!0.027$ and $a\!\simeq\!0.086$, which respectively correspond to the boundary hexagon diameter $L\!\simeq\!13.8$ and $L\!\simeq\!30$, obtained under $\lambda\!=\!0.9$ and $\kappa\!=\!0.65$.

\begin{figure}[hbt]
\centering
\includegraphics[width=8.5cm]{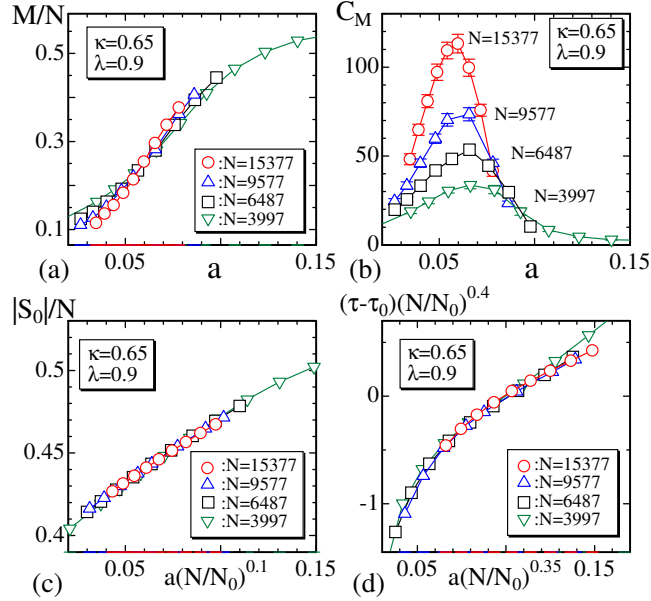}  
\caption{(a) The magnetization $M/N$ vs. the projected triangle area $a$, (b) the variance $C_M$ vs.  $a$, (c)  the absolute internal energy $|S_0|/N$ vs. $a(N/N_0)^{0.1}$, and (d) the frame tension $(\tau\!-\!\tau_0)(N/N_0)^{0.4}$ vs. $a(N/N_0)^{0.35}$, where $\tau_0\!=\!-0.2$ and $N_0\!=\!2437$. }
\label{fig-8}
\end{figure}
The magnetization $M/N$ and its variance $C_M$ are plotted in Figs. \ref{fig-8}(a),(b), and we find that the peak of $C_M$ increases with increasing $N$. This implies that the external force, which fixes the surface boundary, can influence the internal variable $\sigma$. Moreover, we find that $M/N$ scales as a function of $a$ at the transition region, which is shown in Fig. \ref{fig-8}(a). Apparently, this is a nontrivial result, because there is exactly no dependence of $M$ on the projected area in the case of the ordinal sigma model, which is a model of the ferro-magnetic transition. We also expect from Figs. \ref{fig-8}(a),(b) that $\sigma$ is ordered (disordered) in the limit of $a\!\to\! \infty$ ($a\!\to\! 0$). The rotational symmetry of $\sigma$ is spontaneously broken if the surface is expanded to have sufficiently large area. This is in sharp contrast to the original sigma model, in which the symmetry is spontaneously broken only when $\lambda\!\to\! \infty$. The internal energy $S_0/N$ is also seen to scale as a function of $a(N/N_0)^{0.1}$ as shown in  Fig. \ref{fig-8}(c).
   
The frame tension $\tau$ is defined as in \refcite{Ambjorn-NPB1993,WHEATER-JP1994}
\begin{eqnarray}
\label{frame-tension}
\tau = \left(2\langle S_1\rangle -3N \right)/\left( 2A_P\right).
\end{eqnarray}
This relation comes from the scale invariance of the partition function, which is given by  $\partial_\alpha \log Z(\alpha{\bf r})|_{\alpha=1}\!=\!0 $. Indeed, using the fact that $S_2$ is scale independent and  $S_1(\alpha{\bf r})\!=\!\alpha^2 S_1({\bf r})$, and that $\partial_\alpha \log Z(A_P;\alpha{\bf r})|_{\alpha=1}\!=\!\partial_{\alpha} \log Z(\alpha^{-2}A_P;{\bf r})|_{\alpha=1}$ can be written as $-2A_P\partial_{A_P} \log Z(A_P;{\bf r})$, and the assumption that the partition function for the macroscopic membranes is given by $Z_{\rm mac}=\exp\left(-F \right)$ with the free energy $F\!=\!\tau A_P$, we have $\tau$ in Eq.(\ref{frame-tension}) \cite{Ambjorn-NPB1993}. The results in Fig. \ref{fig-8}(d), shown in a linear scale, indicate that $(\tau\!-\!\tau_0)(N/N_0)^{0.4}$ with $\tau_0\!=\!-0.2$ scales as a function of $a(N/N_0)^{0.35}$ at the transition region.

\subsection{String tension of surfaces with two-point boundaries}\label{point-BND}
\begin{figure}[hbt]
\centering
\includegraphics[width=8.5cm]{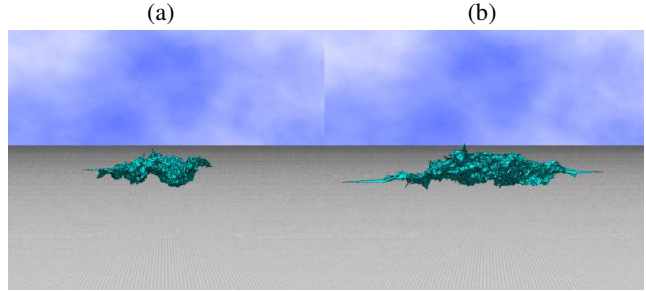}  
\caption{Snapshot of the surfaces of $N\!=\!6487$ for (a) $L\!=\!40$ and (b) $L\!=\!80$, where the parameters are fixed to $\lambda\!=\!0.84$, $\kappa\!=\!0.75$.  The variable $\sigma$ is not shown on the surfaces.}
\label{fig-9}
\end{figure}
The calculation for the string tension measurements needs the boundary condition for the surface, which is  of a hexagon, such that the two boundary vertices are fixed with a sufficiently large separation distance. The two vertices are on a diagonal line of length $L$. This implies that the string tension $\gamma$ is applied to the surface as an external force to make the surface oblong so that the distance between the vertices is fixed to $L$. Snapshots in Figs. \ref{fig-9}(a), (b) obtained with $L\!=\!40$ and $L\!=\!80$ are relatively oblong, although the central parts are not so thin.  

We should note that almost all triangles become oblong on the surface but small part of triangles or vertices, which are located far from the diagonal line between the two terminal points, are influenced only slightly by the extension. For this reason, the scaling behavior of the string tension $\sigma$ with respect to $N$ is expected to deviate from the ordinal one \cite{Ambjorn-NPB1993}.

\begin{figure}[hbt]
\centering
\includegraphics[width=8.5cm]{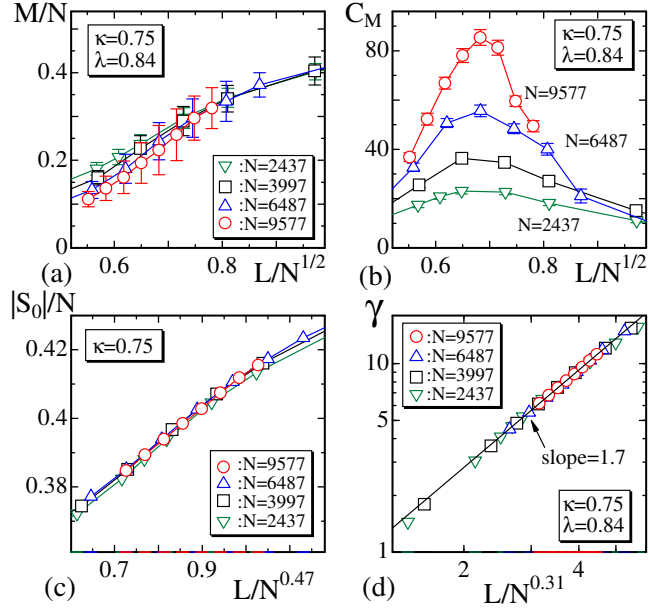}  
\caption{(a) The magnetization $M/N$ vs. $L/\sqrt{N}$, (b) the variance $C_M$ vs. $L/\sqrt{N}$, (c) the internal energy $|S_0|/N$ vs. $L/N^{0.47}$,  and (d) a log-log plot of the string tension $\gamma$ vs. $L/N^{0.31}$, where $\gamma\sim (L/N^{0.31})^{1.7}$ is expected.}
\label{fig-10}
\end{figure}
The magnetization $M/N$ and the variance $C_M$ vs. $L/\sqrt{N}$ are plotted in Figs. \ref{fig-10}(a), (b). The reason why $L/\sqrt{N}$ is used is because $L/\sqrt{N}$ plays a role of the projected chain length $\ell_L$ if the oblong surface is identified with a linear chain of length $L\!=\!\sqrt{N}\ell_L$. This $\ell_L(=\!L/\sqrt{N})$ is expected to be independent of $N$. It corresponds to the projected triangle area $a$ introduced in Section \ref{fixed-BND} for the calculation of the frame tension. The peak of $C_M$ increases with increasing $N$ (Fig. \ref{fig-10}(b)), and this implies that the external force causes an internal transition of $\sigma$ as in the case for the frame tension in Section \ref{fixed-BND}. Thus, the symmetry for $\sigma$ is also spontaneously broken if the surface is expanded to be oblong by external forces.

Now the string tension $\gamma$ is again defined as in \refcite{Ambjorn-NPB1993,WHEATER-JP1994}
\begin{eqnarray}
\label{stringe-tension}
\gamma = \left(2\langle S_1\rangle -3N \right)/\left( 3L\right).
\end{eqnarray}
This formula for  $\gamma$ is obtained by the scale invariance of $Z$ for the surfaces with fixed two point boundaries separated by $L$. 

The internal energy $|S_0|/N$ and the string tension $\gamma$ are expected to scale as a function of $L/N^\alpha$.  Indeed, we find that $|S_0|/N$ and $\gamma$ scale as a function of $L/N^{0.47}$ and  $L/N^{0.31}$, respectively as shown in Figs. \ref{fig-10}(c), (d). We find that the string tension, plotted in a log-log scale in Fig. \ref{fig-10}(d), scales as $\gamma\sim (L/N^{0.31})^{1.7}$ although the range of the variable $L/N^{0.31}$ is not sufficiently large. This scaling behavior deviates from the ordinary one such as $\gamma \sim (L/N)^\delta$ (see \refcite{Ambjorn-NPB1993}). We suppose that the reason of this deviation is that a small number of vertices does not always contribute to the string tension $\tau$ as mentioned above.

\subsection{String tension at high and low $\lambda$ }\label{canonical}
\begin{figure}[hbt]
\centering
\includegraphics[width=8.5cm]{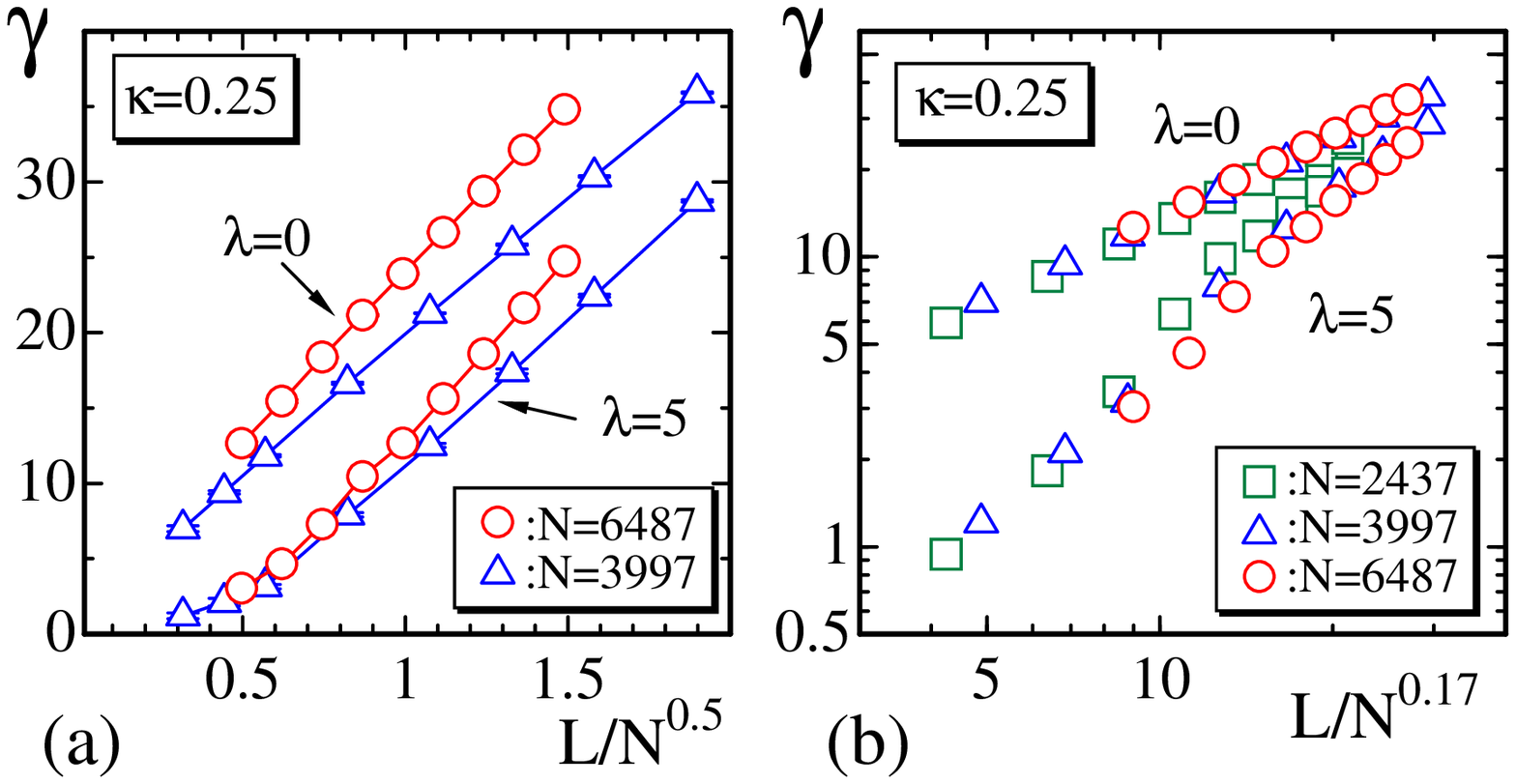}  
\caption{(a) The string tension $\gamma$ vs. $L/\sqrt{N}$ obtained with $\lambda\!=\!0$ and $\lambda\!=\!5$,   and (b) log-log plots for a scaling behavior of $\gamma$, where $\lambda\!=\!0$ and $\lambda\!=\!5$. The bending rigidity is fixed to $\kappa\!=\!0.25$ in both (a) and (b).}
\label{fig-11}
\end{figure}
From the results presented in Figs. \ref{fig-4}(a)--(f), we have found that the surface becomes oblong at sufficiently large $\lambda$ for the small  $\kappa$ region. In this case, the variables $\sigma$ align in the oblong direction. This indicates that the string tension $\gamma$, which is identified with the external force applied to the surface to maintain the surface oblong, is influenced by the presence of the variable $\sigma$ at sufficiently large $\lambda$.

In this subsection, we will show that there is a nontrivial difference between $\gamma$'s at large and small $\lambda$.   
Indeed, $\gamma$ vs. $L/\sqrt{N}$ plotted in Fig. \ref{fig-11}(a) has a large difference in their values, where  $\lambda\!=\!0$ and $\lambda\!=\!5$. The bending rigidity is fixed to a relatively small value $\kappa\!=\!0.25$. The difference of these two $\gamma$'s remains constant and independent of $L/\sqrt{N}$ at least for the region  $0.5\!\leq\!L/\sqrt{N}\!\leq\!1.5$. Therefore, we understand that the difference is quite large compared to  $\gamma$ itself at least for the small $L/\sqrt{N}$ region (see Fig. \ref{fig-11}(a)). 
To see this difference more clearly, we plot $\gamma$ vs. $L/N^{0.17}$ in a log-log scale in Fig. \ref{fig-11}(b). We see that the difference becomes zero in  the sufficiently large  $L/N^{0.17}\!\to\!\infty$. To the contrary, we also see that the difference becomes larger for the smaller $L/N^{0.17}$.  This implies that the surface becomes oblong with small tensile force for the large $\lambda$ region. This phenomenon resembles the soft-elasticity seen in the LCE \cite{Domenici-PNMRS-2012}, and therefore we consider that the Finsler geometric model for membranes can be extended to a model for three dimensional objects such as the LCE. 

\section{Summary and conclusion}\label{conclusion}
We have studied a Finsler geometric model to find how the external mechanical forces influence the internal variable $\sigma$, which represents the tilt and is used to define the Finsler metric. The sigma model interaction energy for the tangential component of $\sigma$ is included into the Hamiltonian with the interaction coefficient $\lambda$.  

Firstly, we find that the model undergoes an anisotropic shape transformation when $\lambda$ changes in the range $0.5\!<\!\lambda\!<\!1.5$ under a fixed bending rigidity $\kappa$ which is relatively small. The surface becomes oblong in the limit of $\lambda\!\to\!\infty$; but it becomes planar and then collapsed as $\lambda$ reduces to $\lambda\!\to\!0$. Secondly, we find that the frame tension $\tau$, as an external force, causes the internal transition of $\sigma$ such that $\sigma$ becomes ordered (disordered) when $\tau$ is sufficiently large (small), which corresponds to large (small) projected triangle area $a$. Thirdly, it is also found that the string tension, as an external force, causes the internal transition of $\sigma$ such that this vector becomes ordered (disordered) when the distance $L$ increases (decreases) between the boundaries. Thus the numerical results presented in this paper confirm that the internal order/disorder transition causes the shape transformations and vice versa in the context of Finsler geometric surface model for membranes. Moreover, the string tension $\gamma$ at the large $\lambda$ region is considerably smaller than  $\gamma$ at  $\lambda\!\to\! 0$. This phenomenon is expected to share the same origin with the soft-elasticity in the liquid crystal elastomer (LCE), and therefore this indicates the possibility that the Finsler geometric model for membranes is extended to a model for $3D$ objects such as the LCE. 

\vspace*{3mm}
\noindent
{\bf Acknowledgments}\\
This work is supported in part by the Grant-in-Aid for Scientific Research (C) Number 26390138. The  author (H.K.) acknowledges the support of Toyohashi University of Technology.




\begin{thebibliography}{00}
\bibitem{GOMPPER-KROLL-SMMS2004}
G. Gompper and D.M. Kroll, in \textit {Statistical Mechanics of Membranes and Surfaces}, Second Edition, edited by  D. Nelson, T. Piran, and S. Weinberg, (World Scientific, 2004) p.359. 

\bibitem{GomKro-PRE1995}
G. Gompper and D.M. Kroll, Phys. Rev. E {\bf 51} 514 (1995).

\bibitem{KWSLipow-PRL1996}
M. Kraus, W. Wintz, U. Seifert, and R. Lipowsky, Phys. Rev. Lett.  {\bf 77}, 3685-3688 (1996).

\bibitem{Nog-Gom-PRL2004}
Hiroshi Noguchi and Gerhard Gompper, Phys. Rev. Lett.  {\bf 93}, 258102 (2004).

\bibitem{KS-PRL2006}
Vasiliy Kantsler and Victor Steinberg, Phys. Rev. Lett.  {\bf 96}, 036001 (2006).

\bibitem{Koibuchi-PRE2007}
Hiroshi Koibuchi, Phys. Rev. E {\bf 76}, 061105(1-5) (2007).

\bibitem{Sefert-etal-PRA1991}
Udo Seifert, Karin Berndl, and Reinhard Lipowsky, Phys. Rev. A {\bf 44}, 1182 (1991).

\bibitem{Sefert-etal-PRE1994}
Ling Miao, Udo Seifert, Michael Wortis, and Hans-G${\ddot {\rm u}}$nther D${\ddot {\rm o}}$bereiner, Phys. Rev. E {\bf 49}, 5389 (1994).

\bibitem{Yanagisawa-Imai-Tani-PRL2008}
Miho Yanagisawa, Masayuki Imai, and Takashi Taniguchi, Phys. Rev. Lett. {\bf 100}, 148102 (2008).

\bibitem{Yanagisawa-Imai-Tani-PRE2010}
Miho Yanagisawa, Masayuki Imai, and Takashi Taniguchi, Phys. Rev. E {\bf 82}, 051928 (2010).

\bibitem{Radzihovsky-SMMS2004}
L. Radzihovsky, in {Statistical Mechanics of Membranes and Surfaces, Second Edition}, edited by  D. Nelson, T. Piran, and S. Weinberg, (World Scientific, 2004) p.275. 

\bibitem{Radzihovsky-Toner-PRL1995}
L. Radzihovsky and J. Toner, Phys. Rev. Lett {\bf 75} 4752 (1995).

\bibitem{Radzihovsky-Toner-PRE1998}
L. Radzihovsky and J. Toner,  Phys. Rev. E {\bf 57} 1832 (1998).

\bibitem{Koibuchi-Sekino-PhysicaA2014}
 H. Koibuchi and H. Sekino, Physica A {\bf 393} 37 (2014).

\bibitem{Domenici-PNMRS-2012}
V. Domenici, Prog. in Nucl. Mag. Res. Spec. {\bf 63}  1 (2012).

\bibitem{HELFRICH-1973}
 W. Helfrich, Z. Naturforsch {\bf 28c} 693 (1973).

\bibitem{POLYAKOV-NPB1986}
 A.M. Polyakov, Nucl. Phys. B {\bf 268} 406 (1986).

\bibitem{FDAVID-SMMS2004}
F. David, in {Statistical Mechanics of Membranes and Surfaces, Second Edition}, edited by  D. Nelson, T. Piran, and S. Weinberg, (World Scientific, 2004) p.149. 

\bibitem{Cai-Lub-PNelson-JFrance1994}
W. Cai, T. C. Lubensky, P. Nelson, and T. Powers, J. Phys. II France
 {\bf 4}, 931 (1994).

\bibitem{Ambjorn-NPB1993}
J. Ambj${\ddot {\rm o}}$rn, A. Irb${\ddot {\rm a}}$ck, J. Jurkiewicz, B. Petersson, Nucl. Phys .B{\bf 393}, Issue 3, 571 (1993). 

\bibitem{WHEATER-JP1994}
J.F. Wheater, J. Phys. A Math. Gen. {\bf 27}, 3323 (1994). 


\bibitem{KOIBUCHI-JPRE2004}
H.Koibuchi, N.Kusano, A.Nidaira, K.Suzuki, and M.Yamada,  Phys. Rev. E {\bf 69} 066139(1-6) (2004). 


\end{thebibliography}
\end{document}